\documentclass[prl,twocolumn,groupedaddress,showpacs]{revtex4}
\usepackage[dvips]{graphics,color}
\usepackage{amsmath,bm,amsfonts}
\DeclareGraphicsExtensions{.eps}
\def\br{{\bm r}}
\def\bC{{\bm C}}\def\bD{{\bm D}}\def\bG{{\bm G}}
\def\bL{{\bm L}}\def\bT{{\bm T}}

\def\cG{{\cal G}}

\begin{document}

\bibliographystyle{revtex}
\title{{\bf Super-slippery Carbon Nanotubes: Symmetry Breaking breaks friction}}
\author{M. Damnjanovi\'c}
\email{yqoq@afrodita.rcub.bg.ac.yu}\homepage{http://www.ff.bg.ac.yu/qmf/qsg_e.htm}
\author{T. Vukovi\'c and I. Milo\v sevi\'c}
\affiliation{Faculty of Physics, University of Belgrade, POB 368,
Belgrade 11001, Yugoslavia}
\date{\today}
%%%%%%%%%%%%%%%%%%%%%%%%%%%%%%%%%%%%%%%%%%%%%%%%%%%%%%%%%%%%%%%%%%%%%%%%%%
\begin{abstract}
The friction between the walls of multi-wall carbon nanotubes is
shown to be extremely low in general, with important details related
to the specific choice of the walls. This is governed by a simple
expression revealing that the phenomenon is a profound consequence of
the specific symmetry breaking: super-slippery sliding of the
incommensurate walls is a Goldstone mode. Three universal principles
of tribology, offering a recipe for the lubricant selection are
emphasized.
\end{abstract}
%%%%%%%%%%%%%%%%%%%%%%%%%%%%%%%%%%%%%%%%%%%%%%%%%%%%%%%%%%%%%%%%%%%%%%%%%%
\pacs{61.48. +c, 62.20.Qp, 64.70.Rh} \maketitle

Various possible applications of the carbon
nanotubes~\cite{IIJ91-SAI98} motivated extensive investigations of
their remarkable physical properties, from the mechanical to the
optoelecronic ones. The symmetry based prediction~\cite{CSYM} of the
extremely small friction between the walls of a double-wall tube
(DWT) has been confirmed recently both by the thorough numerical
analysis~\cite{KOL00} and the experiment~\cite{YU00-CUM00}. Here we
give complete theoretical discussion of such a behavior. To this end
we analyze the DWTs W$'$-W (Tab. \ref{Tdw1212}). The wall W$'$ is
(12,12); its radius is $D'/2=8.14${\AA}. The wall W is one of the
remaining tubes $(n_1,n_2)$. The first three with radii $D/2\approx
R^{\mathrm{in}}=4.7${\AA}, suite as the inner wall, and the last four
with $D/2\approx R^{\mathrm{out}}=11.57${\AA}, as the outer one. The
first part of the paper gives the exhaustive insight to the variety
of the interesting phenomena. Then the results are reconsidered to
extract three general principles profoundly relating interaction and
symmetry, leading to the final conclusion: the reduction of friction
is due to the incompatibility of the DWT walls symmetries.

\begin{figure}[ht]
  \centering
  \includegraphics[1mm,1mm][8cm,8.2cm]{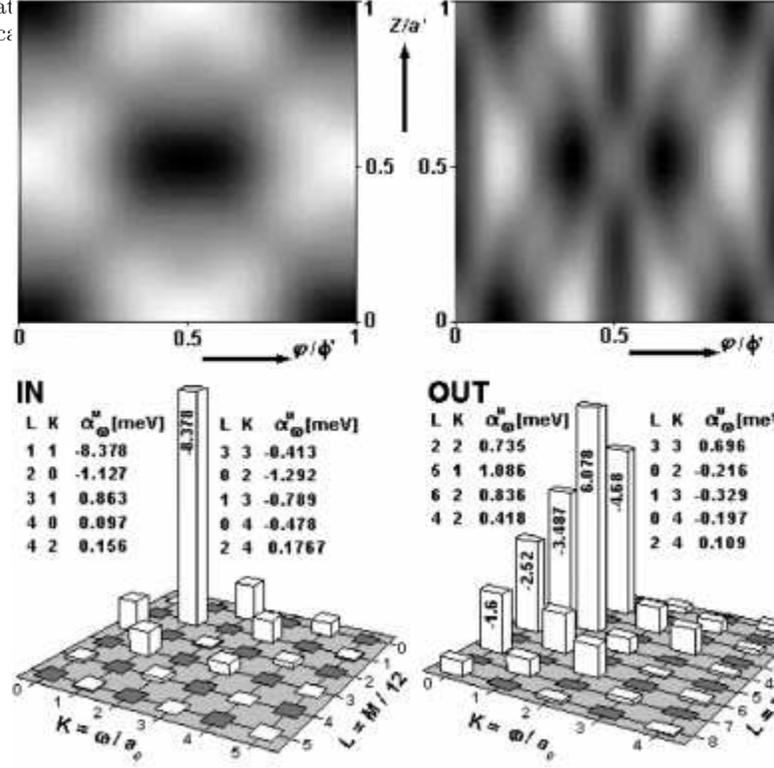}
  \caption{The potential $V(\varphi,z)$ of (12,12) at
  $\rho=R^{\mathrm{in}}$ (left) and $\rho=R^{\mathrm{out}}$ (right)
  within the $\varphi$- and $z$-periods $\phi'$ and $a'$, with the
  expansion amplitudes $\alpha^{M}_\omega$ below.}\label{F12apot}
\end{figure}

Any single-wall tube $(n_1,n_2)$ is completely determined by its
symmetry group~\cite{CSYM}. The $U$-axis maps the atom C$_{000}$ to
the second one C$_{001}$. Then the successive rotations $C_n$ for the
tube $\varphi$-period $\phi=2\pi/n$ ($n=\cG(n_1,n_2)$; $\cG$ -- the
greatest common divisor) yield the initial monomer with $2n$ atoms
C$_{0su}$ ($s=0,\dots,n-1$; $u=0,1$). Finally, the screw axis
generator $(C^r_q|na/q)$ arranges helically the monomers, translating
(for $na/q$) and rotating (through $2\pi r/q$) them along the
$z$-axis. Here, $q$ is the order of the isogonal group principle
axis, $a$ is the translational $z$-period and $r$ is the integer
related to the chirality. Let the $x$-axis of the coordinate system
is to be rotated for $\Phi$ and translated for $Z$ along $z$-axis to
coincide with the tube $U$-axis. Then the coordinates
$\br_{tsu}=(D/2,\varphi_{tsu},z_{tsu})$ of the $t$-th monomer atom
are:
%%%%%%%%%%%%%%%%%%%%%%%%%%%%%%%%%%%%%%%%%%%%%%%%%%%%%%%%%%%%%%%%%%%%%%%
\begin{subequations}\label{Ecoo}\begin{eqnarray}
\varphi_{tsu}&=&(-1)^u\varphi_{0}+2\pi(\frac{tr}{q}+
  \frac{s}{n})+\Phi,%\
% \varphi_0=\frac{n_1+n_2}{\pi D^2/a^2_0}
\label{EcooF}\\
%%%%%%%%%%%%%%%%%%%%%%%%%%%%%%%%%%%%%%%%%%%%%%%%%%%%%%%%%%%%%%%%%%%%%%%%
z_{tsu}&=&(-1)^uz_{0}+t\frac{n}{q}a+Z,\quad t=0,\pm1,\dots .
%  z_0=\frac{n_1-n_2}{\sqrt{2}\pi D}a^2_0
\label{EcooZ}
\end{eqnarray}\end{subequations}
%%%%%%%%%%%%%%%%%%%%%%%%%%%%%%%%%%%%%%%%%%%%%%%%%%%%%%%%%%%%%%%%%%%%%%%
Here $a_0=2.461${\AA}, $ \varphi_0=\frac{n_1+n_2}{\pi D^2}a^2_0$,
$z_0=\frac{n_1-n_2}{\sqrt{2}\pi D}a^2_0$.

Firstly, we look for the potential $V(\varphi,z)$ produced by $W'$ at
the arbitrary point $\br=(D/2,\varphi,z)$. Since necessarily
invariant under the symmetry group of W$'$, $V(\varphi,z)$ can be
expanded in the basis of invariant functions, {\em harmonics}. The
suitable basis invariant under the $U$-axis defined by $\Phi$ and $Z$
is ($M=0,1,\dots$; $\omega$ real):
\begin{subequations}\label{E1Wharm}
\begin{equation}\label{ECharm}
 C^M_\omega(\varphi,z)=
 \cos(M(\varphi-\Phi)+2\pi\omega(z-Z)),
\end{equation}
There is unique combination of $C^{M}_{\omega}$ and
$C^{M}_{-\omega}$, namely
\begin{equation}\label{EAharm}
 A^M_\omega(\varphi,z)=C^{M}_{\omega}(\varphi,z)+
  C^{M}_{-\omega}(\varphi,z),\quad  (M,\omega\geq0),
\end{equation}\end{subequations}
invariant under the mirror planes through $U$. The roto-helical
symmetries impose restrictions on $M$ and $\omega$; indeed, the
invariance under $C_n$ and $(C^r_q|na/q)$ implies:
\begin{subequations}\label{EgenMO}\begin{eqnarray}
 M=0\mod{n},&&\quad\mathrm{(rotations)},\label{EgenMOR}\\
%\omega=0\mod\frac1{a},&&\quad\mathrm{(translations)},\label{EgenMOT}\\
 Mr+\omega na=0\mod q&&\quad\mathrm{(helical)}.\label{EgenMOH}
\end{eqnarray}\end{subequations}
The solutions of this and the other systems appearing latter on will
be given in the form ($L,K=0,\pm1,\dots$)
\begin{equation}\label{Esolgen}
  M=M^*L,\quad \omega=\omega^*K,\quad YL+K=0\mod X,
\end{equation}
with specified $M^*$, $\omega^*$, $X$ and $Y$ ($\cG(X,Y)=1$). In the
rectangular 2D lattice with the periods $M^*$ and $\omega^*$, the
allowed pairs $(M,\omega)$ form the sublattice with the cell $(0,0)$,
$(0,-\omega^*X)$, $(M^*,-\omega^*Y)$ and $(M^*,\omega^*X-\omega^*Y)$.
This cell with area $M^*\omega^*X$ contains only trivial solution
$M=\omega=0$. Particularly, \eqref{EgenMO} is solved by
\eqref{Esolgen} for $\omega^*=a^{-1}$, $M^*=n$, $X=q/n$ and $Y=r$;
the cell area is $\gamma=q/a$. This singles out the harmonics
$C^M_\omega$ and $A^M_\omega$ of the chiral and achiral tubes. The
harmonics $\varphi$- and $z$-periods are $\phi/L$ and $a/|K|$; those
with $M=n$ and $\omega=\pm a^{-1}$ are invariant under the
roto-helical symmetries of W only, while the others have $L|K|$ times
finer periodicity.

 \begin{table}%[h]
  \centering
  \caption{Family of DWTs $(n_1,n_2)$-$(12,12)$.
  The row of W$=(n_1,n_2)$ gives:
  symmetry group of W (line group
   $\bT^r_q(a)\bD_n$ or $\bT^r_q(a)\bD_{nh}$ for W chiral or achiral)
   and of W-W$'$ ($\bG_\cap$), common solutions $X^{M^*}_{\omega}$ of
    \eqref{EgenMO} and \eqref{E12aMO} in the form of \eqref{Esolgen}.
  For commensurate DWT $\bG_\cap=\bT^R_Q(A)\bC_N$ (line group),
   $S_1=\gamma=q/a$, $S_2=\Gamma=Q/A$ and $S=\gamma\gamma'/\Gamma$.
  For incommensurate DWT $\bG_\cap=\bC_N$ (point group),
   $S_1=q$, $S_2=\cG(q,q')$ and $S=qq'\cG(q,q')$. The translational
   periods $a$ and $A$ are in the units of $a_0$.}
  \label{Tdw1212}
\begin{tabular}{ccr|lrrr} \hline\hline
  % after \\ : \hline or \cline{col1-col2} \cline{col3-col4} ...
  Tube&Line Group&$\bG_\cap$&$X^{M^*}_{\omega^*}$&$S_1$&$S_2$&$S$\\ \hline
%%%%%%%%%%%%%%%%%%%%%%%%%%%%%%%%%%%%%%%%%%%%%%%%%%%%%%%%%%%%%%%%%%%%%%%%
  (12,0)&$\bT^{1}_{24}(\sqrt{3})\bD_{12h}$&$\bC_{12}$&$1^{24}_0$&24&24&24\\
%%%%%%%%%%%%%%%%%%%%%%%%%%%%%%%%%%%%%%%%%%%%%%%%%%%%%%%%%%%%%%%%%%%%%%%%
  (7,7)&$\bT^{1}_{14}(1)\bD_{7h}$&$\bT^1_1(1)\bC_1$&$1^{168}_2$&14&1&336\\
%%%%%%%%%%%%%%%%%%%%%%%%%%%%%%%%%%%%%%%%%%%%%%%%%%%%%%%%%%%%%%%%%%%%%%%%
  (11,2)&$\bT^{45}_{98}(7)\bD_{1}$&$\bT^1_1(7)\bC_1$&$7^{168}_2$&14&$\tfrac17$&2352\\
  \hline
%%%%%%%%%%%%%%%%%%%%%%%%%%%%%%%%%%%%%%%%%%%%%%%%%%%%%%%%%%%%%%%%%%%%%%%%
  (12,12)&$\bT^{1}_{24}(1)\bD_{12h}$&&&24&&\\
  \hline
%%%%%%%%%%%%%%%%%%%%%%%%%%%%%%%%%%%%%%%%%%%%%%%%%%%%%%%%%%%%%%%%%%%%%%%%
  (17,17)&$\bT^{1}_{34}(1)\bD_{17h}$&$\bT^1_1(1)\bC_1$&$1^{408}_2$&34&1&816\\
%%%%%%%%%%%%%%%%%%%%%%%%%%%%%%%%%%%%%%%%%%%%%%%%%%%%%%%%%%%%%%%%%%%%%%%%
  (18,16)&$\bT^{51}_{868}(\sqrt{651})\bD_{2}$&$\bC_2$&$1^{5208}_0$&868&4&5208\\
%%%%%%%%%%%%%%%%%%%%%%%%%%%%%%%%%%%%%%%%%%%%%%%%%%%%%%%%%%%%%%%%%%%%%%%%
  (29,1)&$\bT^{1683}_{1742}(\sqrt{2613})\bD_{1}$&$\bC_1$&$1^{20904}_0$&1742&2&20904\\
%%%%%%%%%%%%%%%%%%%%%%%%%%%%%%%%%%%%%%%%%%%%%%%%%%%%%%%%%%%%%%%%%%%%%%%%
  (24,9)&$\bT^{317}_{582}(\sqrt{291})\bD_{3}$&$\bC_{3}$&$1^{2328}_0$&582&6&2328\\
   \hline\hline
\end{tabular}
\end{table}

Consequently, the allowed $M$ and $\omega$ for (12,12) tube,
\begin{equation}\label{E12aMO}
M=12L,\quad\omega=K/a_0,\quad L+K=0\mod2,
\end{equation}
single out the harmonics ${A'}^{M}_{\omega}$ in \eqref{EAharm}.
Further, for the W$'$ wall we fix $\Phi=Z=0$ ($x$-axis is chosen as
the $U$-axis of W$'$). Therefore, one finds:
%%%%%%%%%%%%%%%%%%%%%%%%%%%%%%%%%%%%%%%%%%%%%%%%%%%%%%%%%%%%%%%%%%%%%%%%%%
\begin{equation}\label{EV1}
 V(\varphi,z)=\sum_{tsu}V_C(\br-\br'_{tsu})=
 \sideset{}{'}\sum_{M,\omega\geq0}
 \alpha^{M}_\omega{A'}^{M}_{\omega}(\varphi,z),
\end{equation}
%%%%%%%%%%%%%%%%%%%%%%%%%%%%%%%%%%%%%%%%%%%%%%%%%%%%%%%%%%%%%%%%%%%%%%%%%%
where the prime indicates the summation over the solutions of
\eqref{E12aMO} only. We calculate the potential by the left part of
\eqref{EV1} with $V_C$ being the Van der Waals type interatomic
potential fitted to the $\pi^\bot$-bonding~\cite{KOL00}. The sum over
82 monomers (41 elementary cells with 1968 atoms) is scanned in
$41\times 41$ points within $\varphi\in[0,\pi/6)$ and $z\in[0,1)$.
The fast Fourier transform yields the amplitudes
$\alpha^{M=12L}_{\omega=K}$ ($L,K=0,\dots,20$) on the right of
\eqref{EV1}. The results for $R^{\mathrm{in}}$ and $R^{\mathrm{out}}$
are in Fig. \ref{F12apot}. As the potential $V(\varphi ,z)$
practically does not vary on $10^{-1}$\AA\ scale (in the vicinity of
$R^{\mathrm{in}}$ and $R^{\mathrm{out}}$) these plots refer to all
the DWTs listed in Tab. \ref{Tdw1212}. Arbitrary constant
$\alpha^0_0$ is set to zero. No aliasing occurs, the functions
excluded by \eqref{E12aMO} are with zero amplitudes, while the
amplitudes of the harmonics rapidly decrease with $L$ and $K$, and
become negligible (within the numerical error of $10^{-10}$meV) out
of the depicted range. This property is nicely illustrated in the
density plot (Fig. \ref{F12apot}) of the potential $V(\varphi ,z)$:
It is notably periodic only with the periods of the tube.

With fixed position of W$'$-wall, the W$'$-W relative position and
mutual interaction are determined by $\Phi$ and $Z$ of the W-wall
$U$-axis. The interaction is the sum of the potential \eqref{EV1}
values in the positions \eqref{Ecoo} of W atoms:
%%%%%%%%%%%%%%%%%%%%%%%%%%%%%%%%%%%%%%%%%%%%%%%%%%%%%%%%%%%%%%%%%%%%%%%%%%
\begin{equation}\label{EV2}
 V(\Phi,Z)=
 \sideset{}{'}\sum_{M,\omega\geq0}\sum_{tsu}\alpha^{M}_{\omega}
 A^{'M}_{\omega}(\varphi_{tsu},z_{tsu}).
\end{equation}
%%%%%%%%%%%%%%%%%%%%%%%%%%%%%%%%%%%%%%%%%%%%%%%%%%%%%%%%%%%%%%%%%%%%%%%%%%

For the infinite W-wall the summation over $u$, $s$ and $t$ easily
gives the interaction per the W-wall atom:
%%%%%%%%%%%%%%%%%%%%%%%%%%%%%%%%%%%%%%%%%%%%%%%%%%%%%%%%%%%%%%%%%%%%%%%%%%
\begin{eqnarray}\label{EV2gen}
 v_\infty(\Phi,Z)&=&\sideset{}{''}\sum_{M\geq0,\omega}
  \cos(M\Phi+2\pi\omega Z)\times\nonumber\\
  &\times&2\alpha^{M}_{|\omega|}\cos(M\varphi_{0}+2\pi\omega z_{0}).
  \end{eqnarray}
%%%%%%%%%%%%%%%%%%%%%%%%%%%%%%%%%%%%%%%%%%%%%%%%%%%%%%%%%%%%%%%%%%%%%%%%%%
Here the double prime restricts the summation to $M$ and $\omega$
satisfying simultaneously \eqref{E12aMO} and \eqref{EgenMO} for the
W-wall parameters. This means that the interaction is mediated only
by those W$'$-harmonics which are also invariant under the
roto-helical symmetries of W. Then the allowed $M$-$\omega$ values
(solutions of the system \eqref{EgenMO}-\eqref{E12aMO}) are given by
\eqref{Esolgen} with $Y=1$ and $M^*$, $\omega^*$ and $X$ listed in
Tab. \ref{Tdw1212}. The absolute value of the constants in the second
line of \eqref{EV2gen} roughly determines variation of $v(\Phi,Z)$.
Thus, the corrugation~\cite{KOL00} over W-atom is small whenever the
area $M^*\omega^*X$ is large: large $M$ and $\omega$ indicate small
$|\alpha^{M}_{\omega}|$.

\begin{figure}[ht]
  \centering
  \includegraphics[1mm,1mm][8cm,5.33cm]{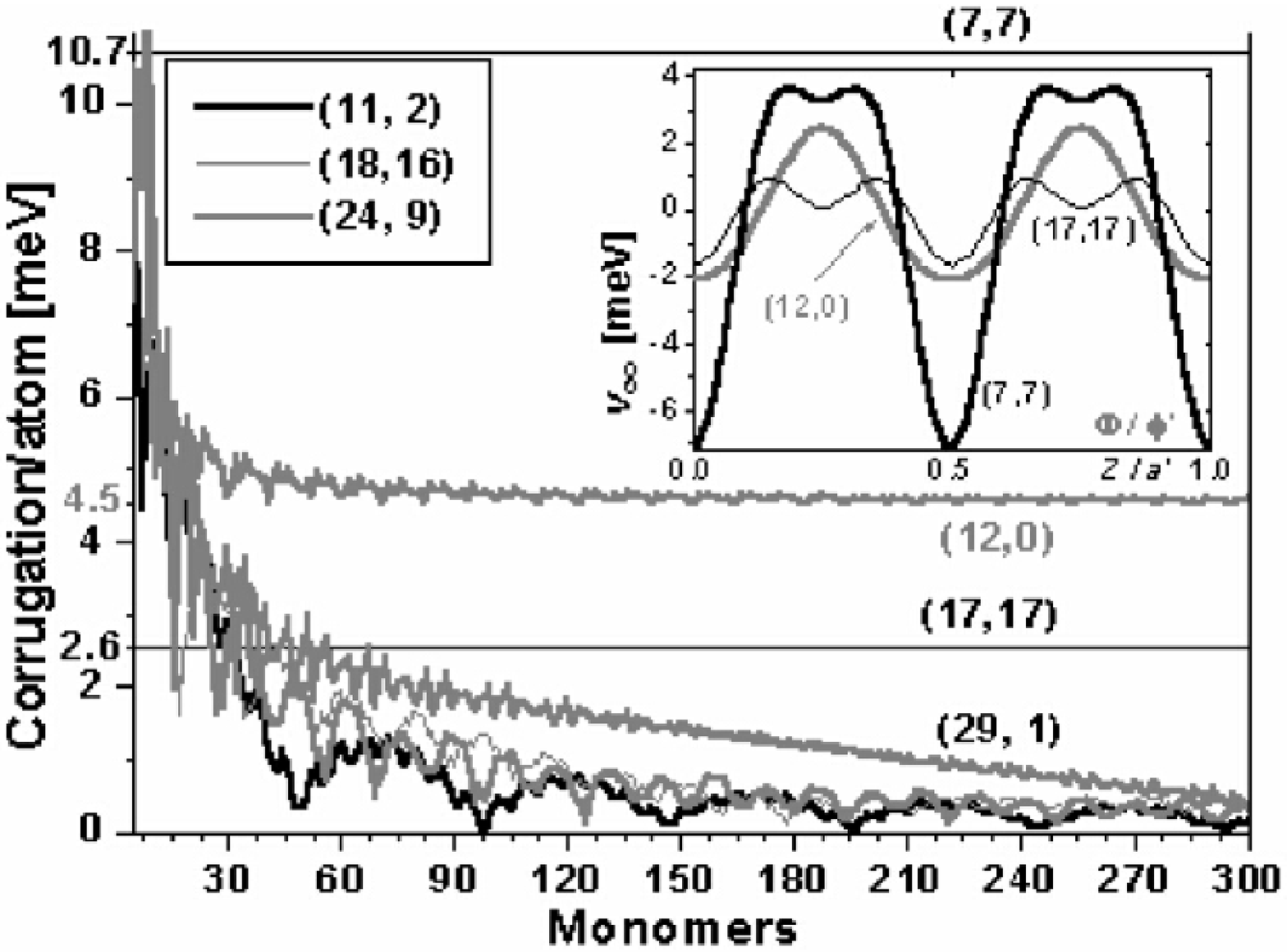}
  \caption{The corrugation/atom dependence of the $W$-wall monomer number
  for the considered DWTs. Inset: $v_\infty(\Phi)$ for the wall
 (12,0), and $v_\infty(Z)$ for (7,7) and (17,17) (for the other walls
 $v_\infty(\Phi,Z)$ is below the precision).}\label{Fcormon}
\end{figure}

For the incommensurate ($a'/a$ irrational) walls $\omega^*=0$.
Consequently the mediating harmonics are constant along $Z$, yielding
$Z$-independent $v_\infty$. Further, $M^*=qq'/\cG(q,q')$ ($X=1$
imposes no restriction). The ideal infinite walls slide along the
tube axis {\em exactly} without friction, while the rotational
friction oscillates with the $\Phi$-period $2\pi/M^*$. Fig.
\ref{Fcormon} presents the calculated interaction $v_\infty(\Phi)$
for the (12,0) wall; its $\Phi$-period is $\phi'/2$. In all other
cases $M^*$ is rather large, allowing only very high $M$ of the
mediators; their amplitudes are beyond the accessible numerical
precision and the computed interaction vanishes, indicating small
friction. In the commensurate cases $M^*$ is large: the
$\Phi$-dependence of the interaction stems from the harmonics with
extremely small amplitudes, giving negligible rotational friction.
The calculated interaction is only $Z$ dependent (Fig.
\ref{Fcormon}), with the $Z$-period $a'/2$ (since $w^*=2$).
Particularly, for the (11,2) wall $X=7$. This additionally forbids
all the numerically significant harmonics: (11,2)-(12,12) walls
friction is extremely weak.

For the finite W-wall with $m$ monomers, the interaction is also
obtained from \eqref{EV2}. The horizontal symmetry axis of W is in
the middle of the ring: $\Phi_m=\Phi+\frac{2\pi r}{q}\frac{m-1}{2}$
and $Z_m=Z+\frac{na}{q}\frac{m-1}{2}$. The summation over $u$, $s$
and $t=0,\dots,m-1$ gives:
%%%%%%%%%%%%%%%%%%%%%%%%%%%%%%%%%%%%%%%%%%%%%%%%%%%%%%%%%%%%%%%%%%%%%%%%%%
\begin{eqnarray}\label{EVmon}
%%%%%%%%%%%%%%%%%%%%%%%%%%%%%%%%%%%%%%%%%%%%%%%%%%%%%%%%%%%%%%%%%%%%%%%%%%
 v_m(\Phi,Z)&=&\sideset{}{''}\sum_{M\geq0,\omega}
 \cos(M\Phi_m+2\pi\omega Z_m)\times\\
 &\times&2\alpha^{M}_{|\omega|}\cos(M\varphi_{0}+2\pi\omega z_{0})
 \frac{\frac{1}{m}\sin(\pi m\frac{rM+na\omega}{q})}
      {\sin(\pi\frac{rM+na\omega}{q})}.\nonumber
      \end{eqnarray}
%%%%%%%%%%%%%%%%%%%%%%%%%%%%%%%%%%%%%%%%%%%%%%%%%%%%%%%%%%%%%%%%%%%%%%%%%%
The double prime here restricts the summation to the solutions of the
system \eqref{EgenMOR}-\eqref{E12aMO}  for $M^*=nn'/N$,
$\omega^*=1/a'$, $X=2$ and $Y=n/N$, with $N=\cG(n,n')$. In fact,
since the ring $W$ has only rotational symmetries, this again
restricts the interaction mediators to the common roto-helical
harmonics. The infinite-wall mediators, satisfying also
\eqref{EgenMOH} are only a part of these.

Analysis of \eqref{EVmon} reveals quite interesting phenomena. The
second line in \eqref{EVmon} apparently suggests $1/m$ dumping of the
corrugation/atom quasi-oscillations. However, both the numerator and
denominator vanish for the infinite walls mediators \eqref{EV2gen},
giving $m$-independent terms which are summed in $v_\infty$. The
additional exclusively finite wall mediators give the dumped
oscillations superposed to $v_\infty$, as transparent from Fig.
\ref{Fcormon}. Although numerically $v_\infty$ may vanish, really it
is only small, and the total corrugation {\em always} increases with
$m$. For the (7,7) wall only the infinite wall mediators are
numerically recorded, while for (17,17) they prevail by several
orders of magnitude. Since they are with $M=0$, the rotational
friction is almost negligible for any $m$. On the contrary, for
(12,0) the rotational friction is dominant. For the other tubes the
infinite wall mediators are not recorded, making both rotational and
translational friction components significantly oscillating with $m$
and approaching to zero. Peculiarity of the commensurate cases is
that whenever $m$ is multiple of $qA/na$, i.e. when the ring length
is multiple of the W-W$'$ translational period $A$, all the
numerators vanish. Only the infinite wall mediators contribute and
$v_m=v_\infty$: the infinite wall corrugation/atom is exactly
regained by this sharp resonant effect! This is illustrated in Fig.
\ref{Fcormon} for (11,2) wall: $v_{98k}$ vanishes and ring moves
coaxially almost freely (if $m$ is an odd multiple of $49$
corrugation is small but observable). For $(7,7)$ and $(17,17)$ this
effect (at $m$ even) is not significant since for any $m$ only the
infinite wall mediators are important.

Giving the complete theoretical insight to the diverse observed
phenomena~\cite{CSYM,KOL00,YU00-CUM00} and predicting some new subtle
effects, the presented analysis itself may be of interest in
tribology. Yet, it can be further deepen by a profound interpretation
based on symmetry, which is a corner stone of the performed harmonic
analysis. To this end, we sublimate the main points of the analysis
into the three general principles. {\em A: the field produced by an
isolated system W$'$ is invariant under the symmetry group of W$'$.}
Otherwise some measurement would identify different symmetry group.
So, W$'$-produced potential $V(\br)$ is invariant function, and can
be expanded over a basis of harmonics only; all but the lowest
harmonics are invariant under supergroups of the W$'$ symmetry group
(finer periods for $M,|\omega|>1$ remarked below \eqref{Esolgen}).
{\em B: the harmonic with large supergroups have small expansion
amplitudes.} Really, $|\alpha^M_\omega|$ sharply decrease and rapidly
become negligible with $M$ and $\omega$. {\em C: the external field
$V$ effects the system W only by the maximal part of $V$ invariant
under the symmetry of $W$.} Thus, only the projection of $V$ to the
subspace of the $W$-harmonics counts, while the contribution of the
orthogonal part vanishes. Indeed, the interaction is mediated only by
those W$'$ harmonics that are additionally invariant under the
roto-helical symmetries of the other wall. Even for them, from
\eqref{E1Wharm} follows that $A^{'M}_{\omega}$ of W$'$ is orthogonal
to all $C^{M'}_{\omega'}$ of $W$ unless $M=M'$ and
$\omega=\pm\omega'$. Therefore the projected $A^{'M}_{\omega}$ is
$C^{M}_{\pm\omega}(\varphi,z)\cos(M\Phi\pm\omega Z)$. Due to the
$W$-wall invariance of $C^{M}_{\omega}$, its value is the same
$C^{M}_{\omega}(\varphi_{0},z_{0})$ over all the W atoms C$_{tsu}$.
This completely explains \eqref{EV2gen}.

Taken together, {\em A} and {\em C} show that the symmetries of the
interacting systems W$'$ and W restrict their interaction only to the
W$'$-harmonics having invariant parts with respect to the W-wall
symmetry. This is performed in two consecutive steps. Firstly, the
roto-helical symmetries of the W and W$'$ walls are collected into
the subgroups ($\bL^{(1)}$-type~\cite{CSYM}) $\bG$ and $\bG'$ of
mutually commuting elements. This enables the choice of the harmonics
of $\bG'$ being either harmonics of $\bG$ (becoming mediators) or
orthogonal to them (giving no contribution). Secondly, as the
parities ($U$-axes) do not commute, they select (for each of the
symmetry groups) the linear combinations \eqref{E1Wharm} of $\bG'$-
and $\bG$-harmonics, which are not mutually orthogonal. The parities
thus influence the interaction by the cosine terms which dump the
mediators amplitudes in \eqref{EV2gen}. Analogously, the first line
in \eqref{EVmon}  is also the dumping cosine, while the second one is
the sum of $C^M_\omega$ values at different monomers (when the ring
symmetry group $\bD_n$ acts on a single atom, a single monomer is
obtained; consequently, the $\bD_n$-harmonic $C^M_\omega$ has
constant value only along a single monomer).

Hence, only the mutually commuting symmetries produce the rarefaction
of the mediators (explicated by the double primes in \eqref{EV2gen}
and \eqref{EVmon}). The corrugation is mainly determined by the
phenomenological harmonic amplitudes (reflecting the nature and
intensity of the interatomic forces), i.e. by the commutative
symmetries of the walls, selecting the corresponding mediators.
Still, all the symmetries contribute to the fine tuning of the
interaction: the dumping factors and the mediators values at the
atoms result in the interaction position dependence. Just this causes
the friction.

Clearly, the rarefaction is the most important effect. Its estimation
in terms of the commuting groups quite profoundly clarifies the
interplay between symmetry and friction. According to the note after
\eqref{Esolgen}, the cell area of the wall harmonic sublattice is
$\gamma=q/a$. This quantity directly measures the roto-helical
symmetry: the more symmetric the wall is, the severer are the
restrictions \eqref{EgenMO}, and the harmonics are sparser with the
larger cell. (Alternatively, since all W-atoms are obtained by the
action of its symmetry group on C$_{000}$, their linear $z$-density
measures the order of the group; and $\gamma$ is the half of this
density.) Further, the area $S=M^*\omega^*X$ points to the lowest
mediators, giving the  estimate of the inverse friction. Simple
inspection of Tab. \ref{Tdw1212} in the commensurate cases shows that
$S=\gamma\gamma'/\Gamma$, where $\Gamma=Q/A$ measures the DWT
symmetry group. The established result in the form
$S=|\bG\otimes\bG'|/|\bG_\cap|$ (the order of the direct product is
$|\bG\otimes\bG'|=|\bG||\bG'|$) means that $S$ is the symmetry
breaking rate from the group $\bG\otimes\bG'$ to the actual DWT
symmetry group $\bG_{\cap}=\bG\cap\bG'$. And, $\bG\otimes\bG'$ is the
symmetry group of the non-interacting systems: independently
performed transformations of $\bG$ and $\bG'$ preserve the system
energy.

As we have shown, the interaction itself imposes a specific symmetry
breaking. In the usual manner this produces
$S=|\bG||\bG'|/|\bG_\cap|$ equivalent minima in the total energy
(recall Jahn-Teller-effect, crystal phase transitions, particle
physics symmetry breaking). While the minima density increases with
the breaking rate, the variation of energy between them, proportional
to the mediators amplitudes, decreases according to the principle
{\em B}. In the extreme cases the energy is constant along some
paths. This Goldstone super-slippery mode of the incommensurate walls
is the phason of the incommensurate phase transitions. Even then the
remaining pure rotational friction reflects the breaking of the
rotational symmetry. The $z$-independent mutual interaction
($\omega=0$) makes the walls effective symmetry groups $\bC_{q'}$ and
$\bC_q$ (isogonal to $\bG'$ and $\bG$), with the intersection
$\bC_{\cG(q,q')}$, and breaking rate $qq'/\cG(q,q')$. For the
incommensurate DWT this equals $M^*$ for \eqref{EV2gen} (Tab.
\ref{Tdw1212}), giving the first allowed mediator $(M^*,0)$. With
$M^*=24$ for (12,0)-(12,12) the walls therefore interact through the
W$'$-harmonic with $L=M/n$'$=2$. Theoretically, this is the lowest
possible DWT mediator, yielding the maximal rotational W-W$'$
friction. To conclude: in the view of the one-to-one correspondence
of the single-wall tubes to their symmetry groups, the symmetries of
the walls are to the large extent incompatible; the tremendous
symmetry breaking and harmonics rarefaction leaves the mediating
harmonics with very small amplitudes.

The performed analysis and conclusions are essentially general, and
refer also to other structures with different type of
symmetries~\cite{FAL00}. For example, for DWT with both walls
incomplete, the corrugation should decrease with the rotational
symmetry breaking rate $nn'/N$. While {\em A} and {\em C} are purely
group theoretical (the proof is based on the group averages), {\em B}
is phenomenological. Its elaboration is beyond the scope of this
letter. Note that only its global validity is expected: the amplitude
of the harmonics may locally increase (e.g.
$|\alpha^{36}_1|>|\alpha^{12}_1|$ for $D/2=R^{\mathrm{out}}$),
revealing the complex structure of the system or some hidden symmetry
of the potential.

Several remarks we add at the end. The minima of the potential
\eqref{EV2gen} determine the equilibrium positions of the W-wall. For
the achiral walls minima of $v_\infty$ (Inset of Fig. \ref{Fcormon})
show that such DWTs have mirror planes~\cite{CSYM}. Linear and
angular momentum quantum numbers of the vibrational modes of the
rotational and translational walls displacements are $k=m=0$; such
modes can be detected by Raman scattering. Their frequencies may be
estimated expanding the potential at minima~\cite{KWO98}. For the
considered sufficiently long DWTs the maximal ones are given by the
Inset of Fig. \ref{Fcormon}: the rotational
$\omega^{(12,0)}_\Phi\approx21$cm$^{-1}$, and two translational
$\omega^{(7,7)}_Z\approx151$cm$^{-1}$,
$\omega^{(17,17)}_Z\approx85$cm$^{-1}$. The others (almost)
frictionless degrees appear as the (almost) acoustic modes. This may
be a hint to test the friction by the Brillouin scattering. Combined
with the described resonant corrugation effect, this can facilitate
the experimental identification of the walls of the sample tubes.
Finally, the rotational and translational displacements are in
general coupled in two helical normal modes, corresponding to the
maximal and minimal gradient of the mediators. This interesting (and
possibly applicable) screwing effect may be estimated within the
presented framework.

%%%%%%%%%%%%%%%%%%%%%%%%%%%%%%%%%%%%%%%%%%%%%%%%%%%%%%%%%%%%%%%%%%%%%%%%%%
\end{document}